\documentclass[a4paper,12pt]{article}
\usepackage{graphicx}
\usepackage{psfrag}

\usepackage{mathrsfs}
\usepackage{amssymb}
\usepackage{amsfonts}
\usepackage{latexsym}
\usepackage{amsmath}
\usepackage{amsthm}
\usepackage[cp1251]{inputenc}
\usepackage[english]{babel}

\newcommand{\er}[1]{\textrm{(\ref{#1})}}
\def\lb{\label}

\theoremstyle{plain}
\newtheorem{Def}{\bf Definition}
\newtheorem{theorem}{\bf Theorem}[section]

\newtheorem{proposition}[theorem]{\bf Proposition}

\theoremstyle{remark}

\renewcommand{\a}{\alpha}

\newcommand{\g}{\gamma}           

\newcommand{\G}{\Gamma}           

\renewcommand{\d}{\delta}           

\newcommand{\D}{\Delta}           \newcommand{\cF}{\mathcal{F}}

\newcommand{\ve}{\varepsilon}     

            \newcommand{\cH}{\mathcal{H}}

\newcommand{\vt}{\vartheta}       

\newcommand{\vT}{\Theta}          

\renewcommand{\k}{\kappa}           

\renewcommand{\l}{\lambda}          \newcommand{\cM}{\mathcal{M}}

\newcommand{\m}{\mu}              

\newcommand{\n}{\nu}              \newcommand{\cP}{\mathcal{P}}

\renewcommand{\r}{\rho}             

\newcommand{\s}{\sigma}           \newcommand{\cR}{\mathcal{R}}

           \newcommand{\cS}{\mathcal{S}}

\renewcommand{\t}{\tau}             \newcommand{\cT}{\mathcal{T}}

\newcommand{\F}{\Phi}             

\newcommand{\vp}{\varphi}

\newcommand{\p}{\psi}             
 
             \newcommand{\cZ}{\mathcal{Z}}
      
\renewcommand{\o}{\omega}
\renewcommand{\O}{\Omega}

\newcommand{\vk}{\varkappa}

  \def\mH{{\mathscr H}}

  \def\mV{{\mathscr V}}

\newcommand{\gD}{\mathfrak{D}}

\newcommand{\gR}{\mathfrak{R}}

\def\Z{\mathbb{Z}}
\def\R{\mathbb{R}}
\def\C{\mathbb{C}}

\def\N{\mathbb{N}}

\def\qqq{\qquad}
\def\qq{\quad}

\let\ge\geqslant
\let\le\leqslant
\let\geq\geqslant

\newcommand{\ca}{\begin{cases}}
\newcommand{\ac}{\end{cases}}
\newcommand{\ma}{\begin{pmatrix}}
\newcommand{\am}{\end{pmatrix}}

\def\lt{\biggl}
\def\rt{\biggr}
\let\geq\geqslant

\renewcommand{\[}{\begin{equation}}
\renewcommand{\]}{\end{equation}}

\def\wt{\widetilde}
\def\pa{\partial}
\def\sm{\setminus}
\def\es{\emptyset}
\def\no{\noindent}

\def\iy{\infty}
\def\ev{\equiv}
\def\/{\over}
\def\ts{\times}

\def\os{\oplus}
\def\ss{\subset}

\def\Re{\mathop{\rm Re}\nolimits}

\def\Im{\mathop{\rm Im}\nolimits}
\def\supp{\mathop{\rm supp}\nolimits}

\def\Tr{\mathop{\rm Tr}\nolimits}

\def\BBox{\hspace{1mm}\vrule height6pt width5.5pt depth0pt \hspace{6pt}}
\def\wh{\widehat}

\begin{document}
\title{Schr\"odinger operators on
armchair nanotubes. I}
\author{
Andrey Badanin
\begin{footnote} {
Arkhangelsk State Technical University,
e-mail: a.badanin@agtu.ru }
\end{footnote}
 \and
 Jochen Br\"uning
 \begin{footnote} {
 Institut f\"ur Mathematik, Humboldt Universit\"at zu Berlin,
 e-mail: bruening@math.hu-berlin.de }
 \end{footnote}
\and
Evgeny Korotyaev
\begin{footnote} {
Correspondence author. Institut f\"ur  Mathematik,  Humboldt Universit\"at zu Berlin,
Rudower Chaussee 25, 12489, Berlin, Germany,
e-mail: evgeny@math.hu-berlin.de\ \
%To whom correspondence should be addressed
}
\end{footnote}
\and Igor Lobanov
\begin{footnote} {
Mathematics Department,
Saint-Petersburg State University of Information Technologies,
       Mechanics and Optics, 1971101, Saint-Petersburg, Sablinskaya 14,
       email: lobanov@mathdep.ifmo.ru
}
\end{footnote}
}

\maketitle

\begin{abstract}
\no We consider the Schr\"odinger operator with a periodic potential on
 quasi-1D models of armchair single-wall nanotubes. The spectrum of this operator consists of an absolutely continuous part (intervals separated by gaps) plus an infinite number of eigenvalues  with infinite multiplicity. We describe all eigenfunctions
with the same eigenvalue. We define a Lyapunov function,
which is analytic on some Riemann surface. On each
sheet, the Lyapunov function has the same properties
as in the scalar case, but it has
branch points, which we call resonances. In example we show the existence
of real and complex resonances for some specific potentials.

\end{abstract}

\section{Introduction  and main results}
\setcounter{equation}{0}

\begin{figure}
\centering
\noindent
\tiny
\hfill
\includegraphics[height=.4\textheight]{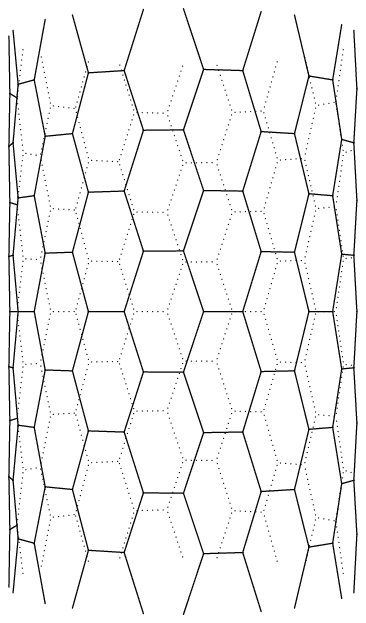}
\hfill
{
\psfrag{g010}{$\Gamma_{0,1}$}
\psfrag{g020}{$\Gamma_{0,2}$}
\psfrag{g030}{$\Gamma_{0,3}$}
\psfrag{g040}{$\Gamma_{0,4}$}
\psfrag{g050}{$\Gamma_{0,5}$}
\psfrag{g060}{$\Gamma_{0,6}$}
\psfrag{g110}{$\Gamma_{1,1}$}
\psfrag{g120}{$\Gamma_{1,2}$}
\psfrag{g130}{$\Gamma_{1,3}$}
\psfrag{g140}{$\Gamma_{1,4}$}
\psfrag{g150}{$\Gamma_{1,5}$}
\psfrag{g160}{$\Gamma_{1,6}$}
\psfrag{g-110}{$\Gamma_{-1,1}$}
\psfrag{g-120}{$\Gamma_{-1,2}$}
\psfrag{g-130}{$\Gamma_{-1,3}$}
\psfrag{g-140}{$\Gamma_{-1,4}$}
\psfrag{g-150}{$\Gamma_{-1,5}$}
\psfrag{g-160}{$\Gamma_{-1,6}$}
\includegraphics[height=.4\textheight]{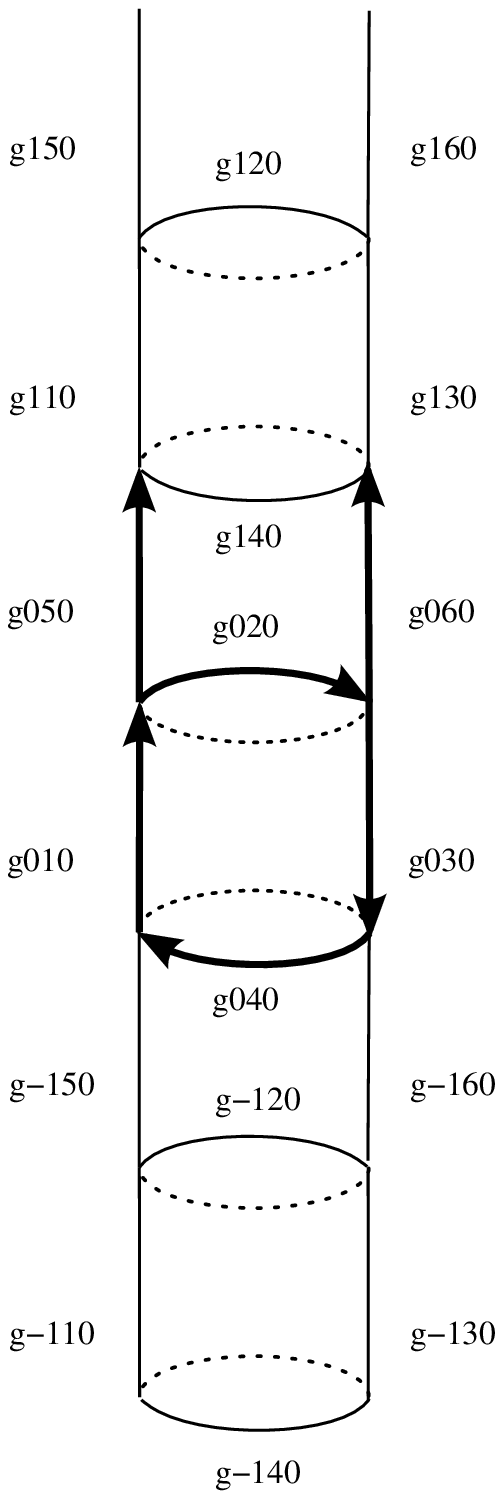}
}
\hfill~
\caption{Armchair graph for $N=10$ and for $N=1$.}
\lb{fig1}
\end{figure}

Consider the Schr\"odinger operator $\mH =-\D+\mV_q$
 with a periodic potential $\mV_q$ on so called armchair
graph $\G^{N},N\ge 1$.
In order to describe the graph $\G^N$, we introduce the fundamental
cell $\wt\G=\cup_{j\in\N_6}\wt\G_j\ss\R^2$, where
$\wt\G_j=\{x=\wt {\bf r}_j+t{\bf e}_j, t\in [0,1]\}$ are edges
of length 1,
\begin{multline}
\N_m=\{1,2,..,m\},\qq
{\bf e}_1={\bf e}_6={1\/2}(1,\sqrt 3),\qq
{\bf e}_2={\bf e}_4=(1,0),\qq {\bf e}_3=-{\bf e}_5={1\/2}(1,-\sqrt 3),\\
\wt{\bf r}_1=(0,0),\qq \wt{\bf r}_2=\wt{\bf r}_5=\wt{\bf r}_1+{\bf e}_1,\qq
\wt{\bf r}_3=\wt{\bf r}_6=\wt{\bf r}_2+{\bf e}_2,\qq
\wt{\bf r}_4=\wt{\bf r}_3+{\bf e}_3.
\end{multline}
We define the strip graph $\wt\G^N$ by
$$
\wt\G^N=\cup_{(n,k)\in\Z\ts\N_N}
(\wt\G+k{\bf e}_h+n{\bf e}_v)\ss\R^2,
\qq {\bf e}_h=(3,0),\qq {\bf e}_v=(0,\sqrt 3).
$$
Vertices of $\wt\G^N$ are given by
$\wt {\bf r}_j+k{\bf e}_h+n{\bf e}_v,(n,j,k)\in\Z\ts\N_6\ts\N_N$.
If we identify the vertices
$\wt{\bf r}_1+n{\bf e}_v$ and $\wt{\bf r}_1+N{\bf e}_h+n{\bf e}_v$
of $\wt\G^N$
for each $n\in\Z$, then we obtain the graph $\G^N$ by
$$
\G^N=\cup_{\o\in \cZ} \G_\o,\qq
\o=(n,j,k)\in \cZ=\Z\ts \N_6\ts \Z_N,
\qq \Z_N=\Z/(N\Z),
$$
where $\G_{\o}=\wt\G_j+k{\bf e}_h+n{\bf e}_v$,
see Fig. \ref{fig1}, \ref{fig2}.
Let ${\bf r}_\o=\wt{\bf r}_j+k{\bf e}_h+n{\bf e}_v$
be a starting point of the edge $\G_{\o}$.
\begin{figure}
\centering
\noindent
{
\tiny
\psfrag{g011}{$\Gamma_{0,1,1}$}
\psfrag{g021}{$\Gamma_{0,2,1}$}
\psfrag{g031}{$\Gamma_{0,3,1}$}
\psfrag{g041}{$\Gamma_{0,4,1}$}
\psfrag{g051}{$\Gamma_{0,5,1}$}
\psfrag{g061}{$\Gamma_{0,6,1}$}
\psfrag{g012}{$\Gamma_{0,1,2}$}
\psfrag{g022}{$\Gamma_{0,2,2}$}
\psfrag{g032}{$\Gamma_{0,3,2}$}
\psfrag{g042}{$\Gamma_{0,4,2}$}
\psfrag{g052}{$\Gamma_{0,5,2}$}
\psfrag{g062}{$\Gamma_{0,6,2}$}
\psfrag{g01N}{$\Gamma_{0,1,N}$}
\psfrag{g02N}{$\Gamma_{0,2,N}$}
\psfrag{g03N}{$\Gamma_{0,3,N}$}
\psfrag{g04N}{$\Gamma_{0,4,N}$}
\psfrag{g05N}{$\Gamma_{0,5,N}$}
\psfrag{g06N}{$\Gamma_{0,6,N}$}
\psfrag{g111}{$\Gamma_{1,1,1}$}
\psfrag{g121}{$\Gamma_{1,2,1}$}
\psfrag{g131}{$\Gamma_{1,3,1}$}
\psfrag{g141}{$\Gamma_{1,4,1}$}
\psfrag{g151}{$\Gamma_{1,5,1}$}
\psfrag{g161}{$\Gamma_{1,6,1}$}
\psfrag{g112}{$\Gamma_{1,1,2}$}
\psfrag{g122}{$\Gamma_{1,2,2}$}
\psfrag{g132}{$\Gamma_{1,3,2}$}
\psfrag{g142}{$\Gamma_{1,4,2}$}
\psfrag{g152}{$\Gamma_{1,5,2}$}
\psfrag{g162}{$\Gamma_{1,6,2}$}
\psfrag{g11N}{$\Gamma_{1,1,N}$}
\psfrag{g2N}{$\Gamma_{1,2,N}$}
\psfrag{g13N}{$\Gamma_{1,3,N}$}
\psfrag{g14N}{$\Gamma_{1,4,N}$}
\psfrag{g15N}{$\Gamma_{1,5,N}$}
\psfrag{g16N}{$\Gamma_{1,6,N}$}
\psfrag{g-111}{$\Gamma_{-1,1,1}$}
\psfrag{g-121}{$\Gamma_{-1,2,1}$}
\psfrag{g-131}{$\Gamma_{-1,3,1}$}
\psfrag{g-141}{$\Gamma_{-1,4,1}$}
\psfrag{g-151}{$\Gamma_{-1,5,1}$}
\psfrag{g-161}{$\Gamma_{-1,6,1}$}
\psfrag{g-112}{$\Gamma_{-1,1,2}$}
\psfrag{g-122}{$\Gamma_{-1,2,2}$}
\psfrag{g-132}{$\Gamma_{-1,3,2}$}
\psfrag{g-142}{$\Gamma_{-1,4,2}$}
\psfrag{g-152}{$\Gamma_{-1,5,2}$}
\psfrag{g-162}{$\Gamma_{-1,6,2}$}
\psfrag{g-11N}{$\Gamma_{-1,1,N}$}
\psfrag{g-12N}{$\Gamma_{-1,2,N}$}
\psfrag{g-13N}{$\Gamma_{-1,3,N}$}
\psfrag{g-14N}{$\Gamma_{-1,4,N}$}
\psfrag{g-15N}{$\Gamma_{-1,5,N}$}
\psfrag{g-16N}{$\Gamma_{-1,6,N}$}
\psfrag{O1}{$\Omega_{1}$}
\psfrag{O2}{$\Omega_{2}$}
\includegraphics[width=.8\textwidth,height=.5\textwidth]{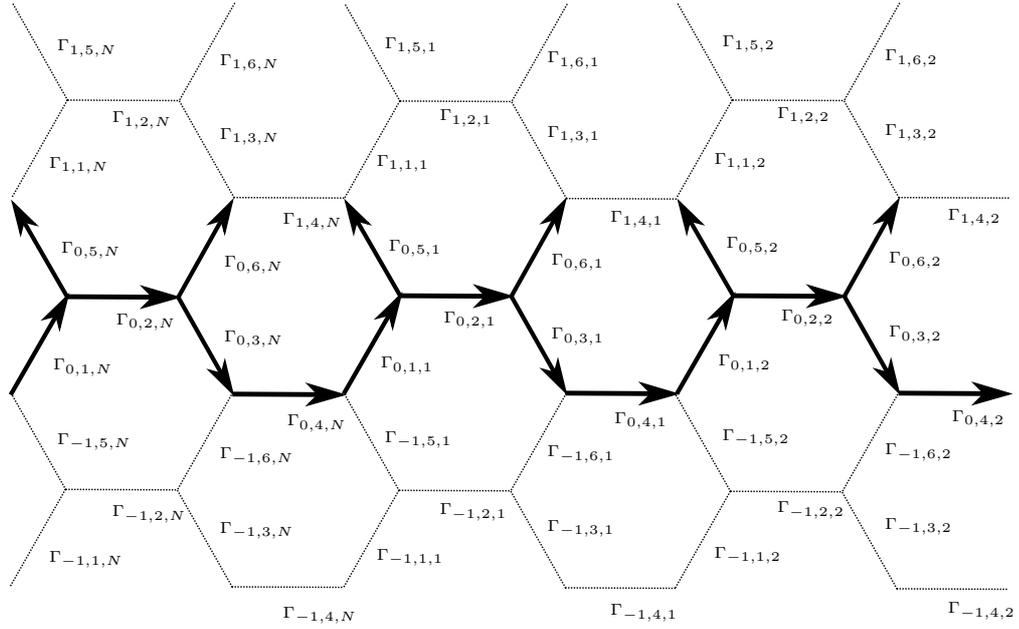}
}
\caption{ A piece of a nanotube $\G^N$.
The fundamental domain is marked by a bold line.
%The fundamental domain $\Gamma_0=\Gamma_{0,0}\cup\Gamma_{0,1}\cup\Gamma_{0,2}$
%is marked by solid line.
}
\lb{fig2}
\end{figure}
We have the coordinate
$x={\bf r}_\o+t{\bf e}_j$
and the local coordinate $t\in [0,1]$ on $\G_\o$.
Thus we give an orientation on the edge.
For each function $y$ on $\G^N$ we define a function
$y_\o=y|_{\G_\o}, \o\in \cZ$.
We identify each function $y_\o$ on $\G_\o$ with a
function on $[0,1]$ by using the local coordinate $t\in [0,1]$.
Define the Hilbert space $L^2(\G^N)=\os_{\o\in \cZ} L^2(\G_\o)$.
Let $C(\G^N)$ be the space   of continuous functions on $\G^N$.
We define the Sobolev space $W^2(\G^N)$ that consists of all
functions $y=(y_\o)_{\o\in\cZ}\in L^2(\G^N),
(y_\o'')_{\o\in\cZ}\in L^2(\G^N)$ and satisfy

\no {\bf   Kirchhoff Boundary Conditions:} {\it $y\in C(\G^N)$
and for each vertex $A$ of $\G^N$ satisfies:
\[
\lb{KirC}
\sum_{\o\in E_A}(-1)^b y_\o'(b)=0,\qq \text{where}\qq
E_A=\{\o\in\cZ:A\in\G_\o\},\qq b=b(\o,A),
\]
where if $A={\bf r}_\o$ is a starting point of $\G_\o$
$($i.e. $t=0$ at $A)$,
then $b(\o,A)=0$,

\qq if $A={\bf r}_\o+{\bf e_j}$ is an endpoint of $\G_\o$
$($i.e. $t=1$ at $A)$,
then $b(\o,A)=1$.
}

The Kirchhoff Conditions \er{KirC} mean  that
the sum of derivatives of $y$ at each vertex of $\G^N$
equals 0 and the orientation of edges gives
the sign $\pm$. Our operator $\mH$ on $\G^N$ acts in the Hilbert space
$L^2(\G^N)$ and is given  by
$(\mH y)_\o=-y_\o''+q y_\o$, where $\ y=(y_\o)_{\o\in \cZ}\in
\gD(\mH)=W^2(\G^N)$  and $(\mV_q y)_\o=qy_\o, q\in L^2(0,1)$.
Note that the orientation of edges is not important for the case
of even potentials
$q\in L_{even}^2(0,1)=\{q\in L^2(0,1):q(t)=q(1-t),t\in(0,1)\}$.
The standard arguments (see \cite{KL}) yield  that $\mH$ is
self-adjoint.

\begin{figure}
\noindent
\centering
\tiny
\psfrag{A}[r][r]{$A_1$}
\psfrag{B}[r][r]{$A_2$}
\psfrag{a}{$\O_1$}
\psfrag{b}{$\O_2$}
\psfrag{c}{$\O$}
\psfrag{(1,0)}{$(1,0)$}
\psfrag{(2,0)}{$(2,0)$}
\psfrag{(3,0)}{$(3,0)$}
\psfrag{(4,0)}{$(4,0)$}
\psfrag{(1,1)}{$(1,1)$}
\psfrag{(2,1)}{$(2,1)$}
\psfrag{(3,1)}{$(3,1)$}
\psfrag{(2,2)}{$(2,2)$}
\psfrag{(3,2)}{$(3,2)$}
\includegraphics[width=.5\textwidth]{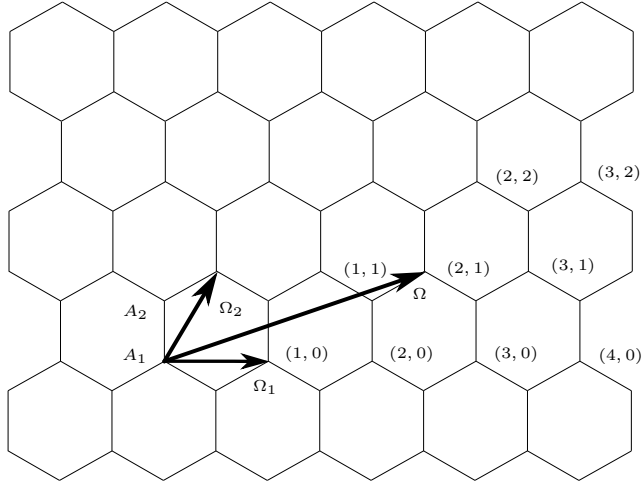}
\caption{The honeycomb lattice of nanotube.
The unit cell is spanned by the vectors $\O_1$ and $\O_2$.
The type of the nanotube is defined by the pair $(N_1,N_2)\in\N^2$,
$N_1\geq N_2$, and corresponding chiral vector
$\O=N_1\O_1+N_2\O_2$.
}
\label{fig5}
\end{figure}

For  the convenience of the reader we  briefly describe the structure
of carbon
nanotubes, see \cite{Ha}, \cite{SDD}. Graphene is a single 2D layer
of graphite
forming a honeycomb lattice, see Fig. \ref{fig5}.
A carbon nanotube is a honeycomb lattice "rolled up" into a
cylinder, see Fig. \ref{fig1}.
In carbon nanotubes, the graphene sheet is "rolled up" in such a way
that the so-called chiral  vector $\O=N_1\O_1+N_2\O_2$ becomes
the circumference of the tube, where $\O_1, \O_2$ are defined in
 Fig \ref{fig5}. The chiral  vector $\O$, which is usually denoted by the pair
of integers $(N_1,N_2)$, uniquely defines a particular tube.
Tubes of type $(N,0)$ are called zigzag tubes.
$(N,N)$-tubes are called armchair tubes.

Our Schr\"odinger operator can be considered
as a model Hamiltonain for $\pi$-electrons in
armchair single-wall carbon nanotubes (SWCN)
which attract tremendous interest from fundamental science and
technological perspectives [AL].
The best known model of SWCN
is a discrete one obtained in the nearest-neighborhood
tight-binding approximation proposed in 1992
(see the articles \cite{HSO}, \cite{MDW}, \cite{SFDD} and review in
\cite{SDD})
shortly after the discovery of carbon nanotubes \cite{Ii}.
Recently, the model was applied to SWCN
(in particular to armchair configuration)
subjected to an electric field \cite{PBRL}.
Being qualitative satisfactory,
the tight binding approximation leads sometimes to
quantitative errors, which are commonly addressed to
curvature effects, see \cite{OHKL}.
However, as was shown in \cite{RMT},
the tight-binding including up to third-nearest
neighbors significantly improves approximation.
To turn on interaction between all atoms in nanotube,
one can proceed to the model considered in \cite{AT},
where carbon atoms was modelled by
zero-range potentials on a cylinder.
In the present article we use
an intermediate model between the mentioned discrete
and two-dimensional ones,
namely we deal with quantum network models
which attract lot of attention over the last decade, see \cite{Ku}.
The considered model was introduced by Pauling \cite{Pa}
and was systematically developed in the series of
articles by Ruedenberg and Scherr \cite{RS}.
Further progress had been made toward periodic systems
by Coulson in \cite{C} where a network model of graphite
layer was worked out.
A network model of a crystal with non-trivial
potentials along bonds was studied by Montroll \cite{M}.
Alexander \cite{A} noticed simple relations
between spectra of quantum graphs
and combinatorial properties of discrete ones
which became common for spectral analysis on
quantum networks at the present time.

The simplest pure carbon nanotube consists of atoms with
covalent bond, when roughly speaking the potential of electric
field between two atoms is even.
%Note that only even potentials were considered in \cite{KuP}.
The situation is more complicated, for example,
in the model of a nitrogen-carbon-boron nanotube
(see \cite{Ha}, Ch.7.3), where
the bond between atoms is ionic and the corresponding potential
is non-even.

Recall the needed properties of the Hill operator $\wt H
y=-y''+q(t)y$ on the real line with a periodic potential
$q(t+1)=q(t),t\in \R$. The spectrum of $\wt H$ is purely absolutely
continuous and consists of intervals
$\wt\s_n=[\wt\l_{n-1}^+,\wt\l_n^-], n\ge 1$. These intervals are
separated by the gaps $\wt\g_n=(\wt\l_n^-,\wt\l_n^+)$ of length
$|\wt\g_n|\ge 0$. If a gap $\wt\g_n$ is degenerate, i.e.
$|\wt\g_n|=0$, then the corresponding segments $\wt\s_n,\wt\s_{n+1}$
merge. For the equation $-y''+q(t)y=\l y$ on the real line we define
the fundamental solutions $\vt(t,\l)$ and $\vp(t,\l),t\in \R$
satisfying $\vt(0,\l)=\vp'(0,\l)=1, \vt'(0,\l)=\vp(0,\l)=0$.
Let
$\vp_1=\vp(1,\cdot),\vt_1=\vt(1,\cdot),\vp_1'=\vp'(1,\cdot),
\vt_1'=\vt'(1,\cdot)$.
The
corresponding monodromy matrix $\wt\cM$, the Lyapunov function $F$,
and the standard function $F_-$ are given by
\[
\wt\cM=\ma\vt_1 & \vp_1 \\
\vt_1' & \vp_1'\am,\qq
F={\vp_1'+\vt_1\/2},\qq  F_-={\vp_1'-\vt_1\/2}.
\]
%The function $F$ has only simple zeros $\e_n,n\ge 1$, which satisfy
%$\e_1<\e_2<...$
The sequence $\wt\l_0^+<\wt\l_1^-\le \wt\l_1^+\ <...$ is the
spectrum of the equation $-y''+qy=\l y$ with two periodic boundary
conditions,  that is  $y(t+2)=y(t), t\in \R$.
 Here equality $\wt\l_n^-= \wt\l_n^+$ means that $\wt\l_n^\pm$ is
 an eigenvalue of
multiplicity 2. Note that $F(\wt\l_{n}^{\pm})=(-1)^n, \  n\ge 1$.
The lowest  eigenvalue $\wt\l_0^+$ is simple, $F(\wt\l_0^+)=1$, and
the corresponding eigenfunction has period 1. The eigenfunctions
corresponding to $\wt\l_n^{\pm}$ have period 1 if $n$ is even, and
they are anti-periodic, that is $y(t+1)=-y(t),\ t\in \R$, if $n$ is
odd. The derivative of the Lyapunov function has a zero $\wt\l_n$ in
each interval $[\l^-_n,\l^+_n]$, that is $ F'(\wt\l_n)=0$.  Let
$\m_n, n\ge 1,$ be the spectrum of the problem $-y''+qy=\l y,
y(0)=y(1)=0$ (the Dirichlet spectrum), and let $\n_n, n\geq 0,$ be
the spectrum of the problem $-y''+qy=\l y, y'(0)=y'(1)=0$ (the
Neumann spectrum). Define the set $\s_D=\{\m_n, n\ge 1\}$ and note
that $\s_D=\{\l\in\C: \vp(1,\l)=0\}$. It is well-known that $\m_n,
\n_n \in [\wt\l^-_n,\wt\l^+_n ],n\ge 1,$ and $\n_0\le \wt\l^+_0$.
%A potential $q$ is even, if $q\in L^2_{even}(0,1)=\rt\{q\in
%L^2(0,1): q(1-t)=q(t),t\in[0,1]\rt\}$.

For simplicity we shall denote $\G_{\a,1}\ss \G^1$ by  $\G_{\a}$, for
$\a=(n,j)\in \cZ_1=\Z\ts \N_6$. Thus $\G^1=\cup_{\a\in \cZ_1} \G_\a$,
see Fig \ref{fig1}. In Theorem \ref{T1} we will show that
 $\mH$ is unitarily equivalent to $H=\os_1^N H_k$, where the operator $H_k$ acts in the Hilbert space $L^2(\G^1)$ and is given by
$(H_k f)_\a=-f_\a''+q f_\a,\ (f_\a)_{\a\in \cZ_1},
(f_\a'')_{\a\in \cZ_1}\in L^2(\G^1)$,
and the components $f_\a,\a\in \cZ_1$  satisfy {\bf the Kirchhoff conditions}:
\begin{multline}
\label{1K0}
f_{n,1}(1)=f_{n,2}(0)=f_{n,5}(0),\qqq
f_{n,2}(1)=f_{n,3}(0)=f_{n,6}(0),\\
f_{n,3}(1)=f_{n,4}(0)=f_{n-1,6}(1),\qq
s^k f_{n,4}(1)=f_{n,1}(0)=f_{n-1,5}(1),\qq s=e^{i{2\pi \/N}},
\end{multline}
\begin{multline}
\label{1K1}
%\label{K12}
f_{n,1}'(1)-f_{n,2}'(0)-f_{n,5}'(0)=0,\qq
f_{n,2}'(1)-f_{n,3}'(0)-f_{n,6}'(0)=0,\\
f_{n,3}'(1)-f_{n,4}'(0)+f_{n-1,6}'(1)=0,\qq
s^kf_{n,4}'(1)-f_{n,1}'(0)+f_{n-1,5}'(1)=0.
\end{multline}

Our analysis is close to previous papers \cite{KL}, \cite{KL1} about the
Schr\"odinger operators on the zigzag graphs. We also introduce the
monodromy matrix and the Lyapunov functions. We study the properties
of these functions and here we essentially use the results and
techniques from the papers \cite{BBK},\cite{CK}, \cite{K1} and
\cite{KL}.

For the operator $H_k$ we define fundamental solutions
$\vt_k^{(\n)}=(\vt_{k,\a}^{(\n)})_{\a\in \cZ_1},
\vp_k^{(\n)}=(\vp_{k,\a}^{(\n)})_{\a\in \cZ_1},\\
 \n=1,2$  which satisfy
\begin{multline}
\lb{eqf0}
-\D y+\mV_qy=\l y, \qq  on \ \ \G^1,\qqq
\text{the Kirchhoff Boundary Conditions \er{1K0},\er{1K1}}\\
\vT_{k,-1}(1,\l)=\F_{k,-1}'(1,\l)=I_2,\qq \vT_{k,-1}'(1,\l)=\F_{k,-1}(1,\l)=0,
\end{multline}
where $I_n,n\ge 2$ is the $n\ts n$ identity matrix and
\[
\lb{dmm1}
\vT_{k,n}=\ma\vt_{k,n,5}^{(1)} & \vt_{k,n,5}^{(2)} \\
\vt_{k,n,6}^{(1)} & \vt_{k,n,6}^{(2)}\am,\qq
\F_{k,n}=\ma\vp_{k,n,5}^{(1)} & \vp_{k,n,5}^{(2)} \\
\vp_{k,n,6}^{(1)} & \vp_{k,n,6}^{(2)}\am.
\]
We define the monodromy matrix (determined by the fundamental
solutions on $\G_0^1$) by
\[
\lb{dmm}
\cM_k(\l)=\ma \vT_{k,0} & \F_{k,0}\\
\vT_{k,0}' & \F_{k,0}'\am(1,\l).
\]
We formulate our first result about the monodromy matrix.

\begin{theorem}
\label{T1}
i) The operator $\mH$ is unitarily equivalent to
$H=\os_1^N H_k$.

\no ii) Let $k\in \Z_N$. Then for any $\l\in\C\sm\s_D$ there exist unique fundamental solutions $\vt_k^{(\n)}=(\vt_{k,\a}^{(\n)})_{\a\in \cZ_1}, \vp_k^{(\n)}=(\vp_{k,\a}^{(\n)})_{\a\in \cZ_1},\n=1,2$.
 Moreover, each of the functions $\vt_{k,\a}^{(\n)}(x,\l)$,
$\vp_{k,\a}^{(\n)}(x,\l)$, $x\in \G^1$ is
analytic in $\l\in\C\sm\s_D$ and the monodromy matrix $\cM_k(\l)$ satisfies
\[
\label{T1-2}
\det \cM_k=1,
\quad
\Tr\cM_0
=2(9F^2-F_-^2-1),\quad
\Tr\cM_k=\Tr\cM_0-4s_k^2,\qq
s_k=\sin \frac{\pi k}{N},
\]
\[
\lb{T1-4}
\Tr\cM_0^2
=72F^2+{1\/2}(\Tr\cM_0)^2-4,\qq
\Tr\cM_k^2=\Tr\cM_0^2-8s_k^2\Tr\cM_0-4s_{2k}^2,
\]
\[
\lb{T1-3}
\cM_k(\l)^\top  J\cM_k(\l)=J,\qqq where \qq
J=\ma 0& {\bf j}_2\\ -{\bf j}_2& 0\am,\ \
{\bf j}_2=\ma 0& 1\\ 1& 0\am,
\]
and the matrix-valued function
$\cR\cM_k\cR^{-1}$ is entire, where $\cR=I_2\os \vp_1I_2$.
\end{theorem}

The monodromy matrix $\cM_k$ has poles at the points $\l\in \s_D$,
which are eigenvalues of $H_k$ (see Theorem \ref{T2}), similar to
the zigzag tube \cite{KL}, \cite{MV}.

Define the subspace $\cH_k(\l)=\{\p\in \gD(H_k): H_k\p=\l \p\}$
for $(\l,k)\in \s_{pp}(H_k)\ts\Z_N$. If
$\dim \cH_k(\l_0)=\iy$ for some $\l_0\in \s_{pp}(H_k)$, then we say that
$\{\l_0\}$ is a flat band.
In Theorem \ref{T2} we describe all flat bands
and corresponding eigenfunctions (see Fig. \ref{fig3}).

\begin{figure}
\centering
\noindent
\tiny
\hfill
\psfrag{g010}{$\Gamma_{0,1}$}
\psfrag{g020}{$\Gamma_{0,2}$}
\psfrag{g030}{$\Gamma_{0,3}$}
\psfrag{g040}{$\Gamma_{0,4}$}
\psfrag{g050}{$\Gamma_{0,5}$}
\psfrag{g060}{$\Gamma_{0,6}$}
\psfrag{g110}{$\Gamma_{1,1}$}
\psfrag{g120}{$\Gamma_{1,2}$}
\psfrag{g130}{$\Gamma_{1,3}$}
\psfrag{g140}{$\Gamma_{1,4}$}
\psfrag{g150}{$\Gamma_{1,5}$}
\psfrag{g160}{$\Gamma_{1,6}$}
\psfrag{g-110}{$\Gamma_{-1,1}$}
\psfrag{g-120}{$\Gamma_{-1,2}$}
\psfrag{g-130}{$\Gamma_{-1,3}$}
\psfrag{g-140}{$\Gamma_{-1,4}$}
\psfrag{g-150}{$\Gamma_{-1,5}$}
\psfrag{g-160}{$\Gamma_{-1,6}$}
a)
\includegraphics[height=.4\textheight]{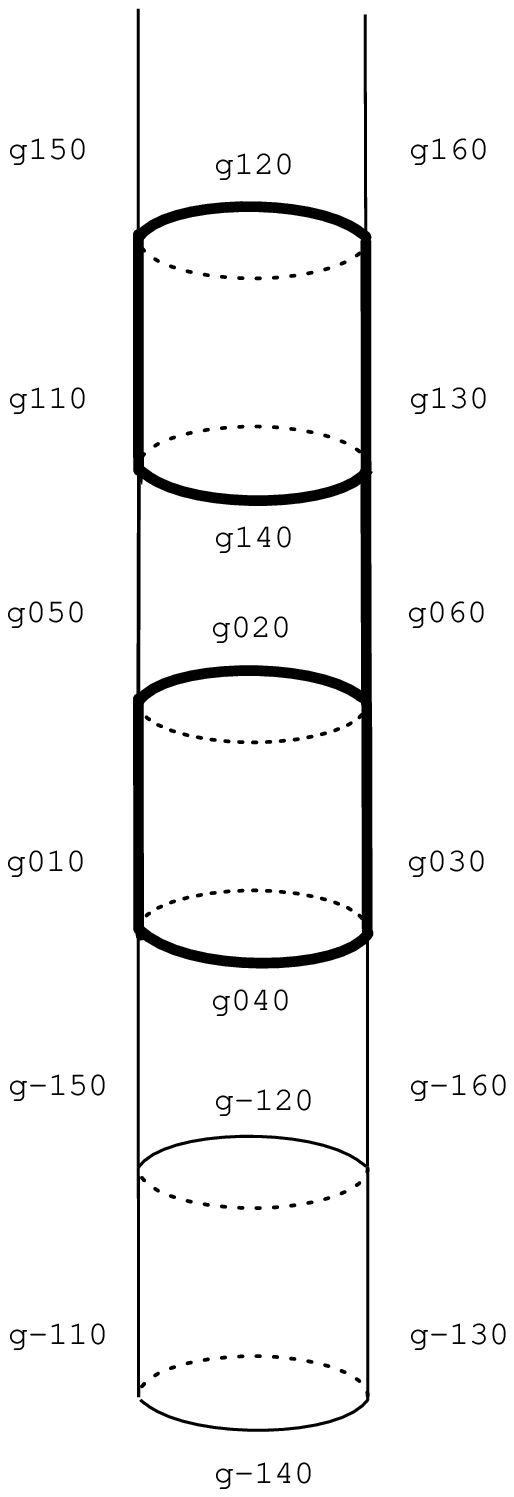}
\hfill
b)
\includegraphics[height=.4\textheight]{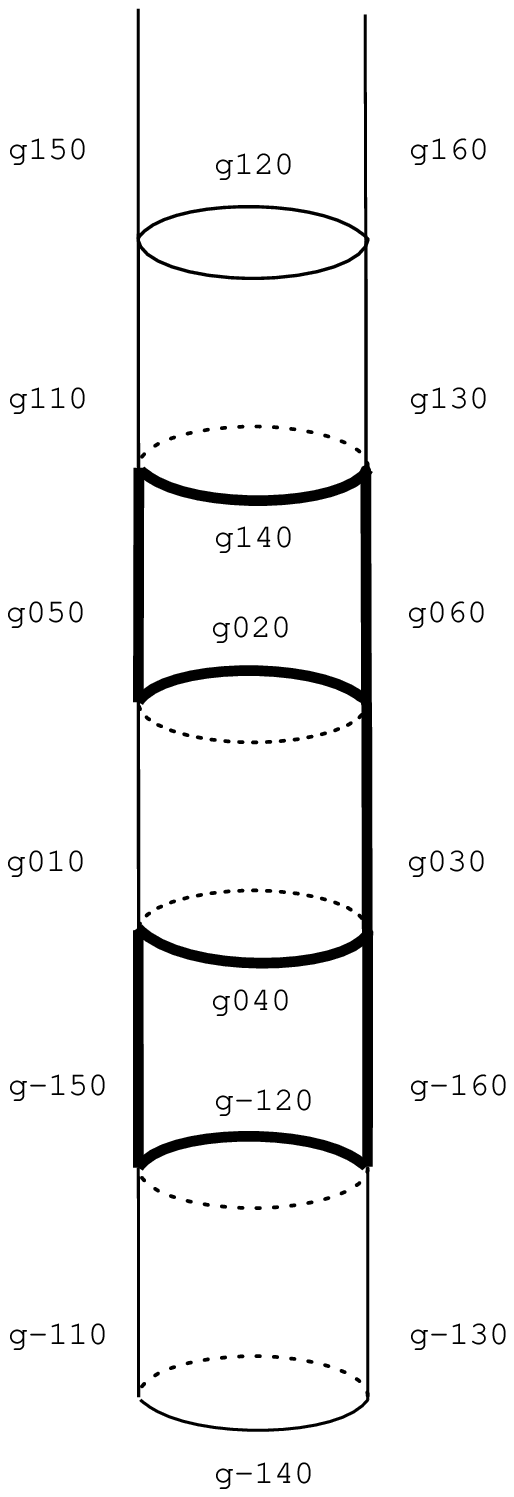}
\hfill~
\caption{The supports of eigenfunctions: a) $\psi^{(0,1)}$,
b) $\psi^{(0,2)}$.}
\lb{fig3}
\end{figure}

\begin{theorem} \lb{T2}
Let $(\l,k)\in\s_D\ts \Z_N$. Then

\no i) Every eigenfunction of $\cH_k(\l)$ vanishes at all verteces
of $\G^1$.

\no ii) There exist two functions
$\p^{(0,\n)}=(\p^{(0,\n)}_\a)_{\a\in
\cZ_1}\in\cH_k(\l),\n=1,2,$ on $\G^1$ such that
$\supp \p^{(0,\n)}\ss \cup_{j=1}^6\G_{0,j}$, each function
$\p^{(n,\n)}=(\p^{(0,\n)}_{n-m,j})_{(m,j)\in \cZ_1}\in \cH_k(\l),
n\in \Z$.
Moreover, each $f\in \cH_k(\l)$ satisfies:
\[
\lb{T2-3}
f=\sum_{n\in\Z} (\wh  f_{n,1}\p^{(n,1)}+\wh  f_{n,2}\p^{(n,2)}),\qqq
(\wh f_{n,1},\wh f_{n,2})_{n\in\Z}\in \ell^2\os\ell^2,
\]
where
$$
\wh f_{n,1}=-\vp_1' f_{n,1}'(0),\ \
\wh  f_{n,2}=f_{n,2}'(0),\qq \text{if} \qq \vp_1'^2=1,k=0,
$$
\[
\lb{T2-4}
\wh  f_{n,1}={\vk_1 f_{n,6}'(0)+ f_{n,3}'(0)\/\vk_1\vk_2-1},
\ \ \wh  f_{n,2}={f_{n,6}'(0)+\vk_2 f_{n,3}'(0)\/\vk_1\vk_2-1},\qq
\text{if} \ \vp_1'^2\ne
1\ \text{or}\ \vp_1'^2=1,k\ne 0,
\]
$\vk_1=1-s^k\vp_1'^2,\vk_2=1-s^k\vp_1'^4.$
Moreover, the mapping $f\to (\wh  f_{n,1},\wh f_{n,2})_{n\in\Z}$
is a linear isomorphism between $\cH_k(\l)$ and $\ell^2\os\ell^2$.

\end{theorem}

Using \er{T1-3} we deduce that eigenvalues of $\cM_k, k\in \Z_N$
have the form $\t_{1,\nu}^{\pm 1},\t_{2,\nu}^{\pm 1}$.

\begin{theorem}
\lb{T3}  Let $\t_{k,1}^{\pm 1 }, \t_{k,2}^{\pm 1 }$
be eigenvalues of $\cM_k$ for some $k\in \Z_N$. Then

\no (i) The Lyapunov functions
$F_{k,\nu}={1\/2}(\t_{k,\nu}+{1\/\t_{k,\nu}}), \n=1,2$ satisfy
\[
\lb{DeLk}
F_{k,\nu}=\xi_k-(-1)^\n\sqrt{\r_k},\qq
\xi_k={9F^2-F_-^2-1\/2}-s_k^2,\qq
\r_k=(9F^2-s_k^2)c_k^2+s_k^2F_-^2,
\]
\[
\lb{S3b} D_k(\t,\l)\ev \det(\cM_k(\l)-\t I_4)
=(\t^2-2F_{k,1}(\l)\t+1)(\t^2-2F_{k,2}(\l)\t+1),
\]
where
%$c_k=\cos \f_k, \ s_k=\sin \f_k$ and
$F_{k,\nu}$ are branches of $F_k=\xi_k+\sqrt{\r_k}$ on a
two sheeted Riemann surface $\gR_k$ defined by $\sqrt {\r_k}$.

\no (ii) If $F_k(\l)\in (-1,1)$ for some $\l\in \R$ and
$\l$ is not a branch point of $F_k$, then $F_k'(\l)\neq 0$.

\no (iii)  The  following identities hold:
\begin{multline}
\lb{T3-1}
\s(H_k)=\s_{\iy}(H_k)\cup\s_{ac}(H_k),\qq  \s_{\iy}(H_k)=\s_D,\\
\s_{ac}(H_k)=\{\l\in\R: F_{k,\nu}(\l)\in [-1,1]\ \text{for some}\
\nu\in\N_2\}.
\end{multline}

\no (iv) If $q\in L_{even}^2(0,1)$,
then $\s_{ac}(H)=\s_{ac}(H_0)=\s(\wt H)$.

\end{theorem}

\begin{Def}
A zero of $\r_k, k\in \Z_N$ is called a {\bf resonance} of $H_k$.
\end{Def}

Let  $r_{k,n}^\pm,n\ge 1$ be zeros of $\r_k$ and let their labeling be
given  by
$\Re r_{k,1}^-\le \Re r_{k,1}^+\le \Re r_{k,2}^-\le \Re r_{k,2}^+\le ...$
We show the existence of real and non-real resonances.

\begin{proposition}
\lb{P7}
Let  $q=q_\ve={1\/v}\d(t-{1\/2}-c_k \ve-\ve^2),\ve\ne 0, t\in [0,1]$
for some $k\not\in\{0,{N\/2}\}$,  and let $n_0>1$.
Then for each $0<n\le n_0$
there exist functions $r_{k,n}^\pm(z^2)$, analytic in the disk
$\{|z|<\ve_n\}$ with some $\ve_n>0$,
such that  $\r_k(r_{k,n}^\pm(\ve),q_\ve)=0$ for all $\ve\in (-\ve_n,\ve_n)\sm\{0\}$
and
\[
\lb{P4-1}
r_{k,n}^\pm(0)=(\pi n)^2,\qqq
r_{k,n}^\pm(\ve)=(\pi n)^2-2\pi n \ve\pm i{4\sqrt{2}(\pi
n)^2s_k\/3\sqrt{c_k}}\ve^{3/2}+O(\ve^{2})\qq  as \ \ v\to 0.
\]
Moreover, $\o_{k,n}^-(\ve)=(r_{k,n}^-(\ve),r_{k,n}^+(\ve))\ss \R$ and
$\r_k(\l,q_\ve)<0$ for all $\l\in \o_{k,n}^-(\ve), \ve\in (-\ve_n,0)$.
\end{proposition}

\no {\it Remark.}
(i) Graph of the function $F_{k}, k\ne \{0,{N\/2}\}$
for $q=q_\ve,\ve\in(0,\ve_1)$ is given by Fig.\ref{lfd}.

\no (ii) The potential $q=q_\ve={1\/v}\d(t-{1\/2}-c_k \ve-\ve^2),v\ne 0$
in Proposition \ref{P7} is not even.

\no (iii) If $\ve\in (-\ve_n,0)$, then $F_{k,\n}(\l)\notin \R$ for all
$\l\in \o_{k,n}^-(\ve)$, which yields $\o_{k,n}^-(\ve)\cap \s(H_k)=\es.$

%%%%%%%%%%%
%%%%%%%%%%
\begin{figure}
\unitlength 1.00mm \linethickness{0.2pt}
\begin{picture}(162.00,60.00)(00.00,-15.00)
%%%%%%%%%%% coordinate lines
\put(10.00,-5.00){\line(0,1){70.00}}
\put(10.00,40.00){\line(1,0){150.00}}
\put(05.00,20.00){\line(1,0){155.00}}
\put(10.00,00.00){\line(1,0){150.00}}
\put(160.00,23.00){\makebox(0,0)[cc]{$\l$}}
\put(05.00,00.00){\makebox(0,0)[cc]{$-1$}}
\put(05.00,40.00){\makebox(0,0)[cc]{$1$}}
\put(17.00,60.00){\makebox(0,0)[cc]{$F_k(\l)$}}
%%%%%%%%%%%% half infinite part
\bezier{600}(10.00,50.00)(30.00,-8.00)(60.00,-10.00)
\bezier{600}(60.00,-5.00)(50.00,-4.00)(25.00,50.00)
%%%%%%%%%%% first boomerang: upper bound
\bezier{400}(60.00,-5.00)(70.00,-3.00)(85.00,40.00)
\bezier{400}(85.00,40.00)(100.00,80.00)(120.00,40.00)
\bezier{400}(120.00,40.00)(140.00,02.00)(130.00,-2.00)
%%%%%%%%%%% first boomerang: lower bound
\bezier{400}(60.00,-10.00)(85.00,-8.00)(97.00,40.00)
\bezier{150}(97.00,40.00)(100.00,47.00)(105.00,40.00)
\bezier{400}(130.00,-2.00)(120.00,02.00)(105.00,40.00)
%%%%%%%%%%% second boomerang: upper bound
\bezier{400}(150.00,-1.00)(140.00,02.00)(161.00,41.00)
%%%%%%%%%%% second boomerang: lower bound
\bezier{250}(150.00,-1.00)(155.00,01.00)(160.00,08.00)
%%%%%%%%%% points
\put(133.50,19.00){\line(0,1){2.00}}
\put(146.70,19.00){\line(0,1){2.00}}
%\put(15.00,15.00){\makebox(0,0)[cc]{$E_{2,0}^{k,+}$}}
\put(15.00,19.90){\line(1,0){22.00}}
\put(15.00,19.80){\line(1,0){22.00}}
\put(15.00,19.70){\line(1,0){22.00}}
\put(15.00,19.60){\line(1,0){22.00}}
\put(15.00,19.50){\line(1,0){22.00}}
\put(15.00,19.40){\line(1,0){22.00}}
\put(15.00,19.30){\line(1,0){22.00}}
\put(15.00,19.20){\line(1,0){22.00}}
\put(30.00,20.10){\line(1,0){22.00}}
\put(30.00,20.20){\line(1,0){22.00}}
\put(30.00,20.30){\line(1,0){22.00}}
\put(30.00,20.40){\line(1,0){22.00}}
\put(30.00,20.50){\line(1,0){22.00}}
\put(30.00,20.60){\line(1,0){22.00}}
\put(30.00,20.70){\line(1,0){22.00}}
\put(30.00,20.80){\line(1,0){22.00}}
\put(67.00,20.10){\line(1,0){18.00}}
\put(67.00,20.20){\line(1,0){18.00}}
\put(67.00,20.30){\line(1,0){18.00}}
\put(67.00,20.40){\line(1,0){18.00}}
\put(67.00,20.50){\line(1,0){18.00}}
\put(67.00,20.60){\line(1,0){18.00}}
\put(67.00,20.70){\line(1,0){18.00}}
\put(67.00,20.80){\line(1,0){18.00}}
\put(79.00,19.90){\line(1,0){17.00}}
\put(79.00,19.80){\line(1,0){17.00}}
\put(79.00,19.70){\line(1,0){17.00}}
\put(79.00,19.60){\line(1,0){17.00}}
\put(79.00,19.50){\line(1,0){17.00}}
\put(79.00,19.40){\line(1,0){17.00}}
\put(79.00,19.30){\line(1,0){17.00}}
\put(79.00,19.20){\line(1,0){17.00}}
\put(105.00,19.90){\line(1,0){19.00}}
\put(105.00,19.80){\line(1,0){19.00}}
\put(105.00,19.70){\line(1,0){19.00}}
\put(105.00,19.60){\line(1,0){19.00}}
\put(105.00,19.50){\line(1,0){19.00}}
\put(105.00,19.40){\line(1,0){19.00}}
\put(105.00,19.30){\line(1,0){19.00}}
\put(105.00,19.20){\line(1,0){19.00}}
\put(120.00,20.10){\line(1,0){13.50}}
\put(120.00,20.20){\line(1,0){13.50}}
\put(120.00,20.30){\line(1,0){13.50}}
\put(120.00,20.40){\line(1,0){13.50}}
\put(120.00,20.50){\line(1,0){13.50}}
\put(120.00,20.60){\line(1,0){13.50}}
\put(120.00,20.70){\line(1,0){13.50}}
\put(120.00,20.80){\line(1,0){13.50}}
\put(130.00,19.90){\line(1,0){3.50}}
\put(130.00,19.80){\line(1,0){3.50}}
\put(130.00,19.70){\line(1,0){3.50}}
\put(130.00,19.60){\line(1,0){3.50}}
\put(130.00,19.50){\line(1,0){3.50}}
\put(130.00,19.40){\line(1,0){3.50}}
\put(130.00,19.30){\line(1,0){3.50}}
\put(130.00,19.20){\line(1,0){3.50}}
\put(147.00,20.10){\line(1,0){13.00}}
\put(147.00,20.20){\line(1,0){13.00}}
\put(147.00,20.30){\line(1,0){13.00}}
\put(147.00,20.40){\line(1,0){13.00}}
\put(147.00,20.50){\line(1,0){13.00}}
\put(147.00,20.60){\line(1,0){13.00}}
\put(147.00,20.70){\line(1,0){13.00}}
\put(147.00,20.80){\line(1,0){13.00}}
\put(147.00,19.90){\line(1,0){2.00}}
\put(147.00,19.80){\line(1,0){2.00}}
\put(147.00,19.70){\line(1,0){2.00}}
\put(147.00,19.60){\line(1,0){2.00}}
\put(147.00,19.50){\line(1,0){2.00}}
\put(147.00,19.40){\line(1,0){2.00}}
\put(147.00,19.30){\line(1,0){2.00}}
\put(147.00,19.20){\line(1,0){2.00}}
\put(153.00,19.90){\line(1,0){7.00}}
\put(153.00,19.80){\line(1,0){7.00}}
\put(153.00,19.60){\line(1,0){7.00}}
\put(153.00,19.50){\line(1,0){7.00}}
\put(153.00,19.40){\line(1,0){7.00}}
\put(153.00,19.30){\line(1,0){7.00}}
\put(153.00,19.20){\line(1,0){7.00}}
\put(153.00,19.70){\line(1,0){7.00}}
%%%%%%%%%%%%%%%%%%%%%%%
\put(15.00,15.00){\makebox(0,0)[cc]{$E_{2,0}^{k,+}$}}
\put(30.00,24.00){\makebox(0,0)[cc]{$E_{1,0}^{k,+}$}}
\put(85.00,24.00){\makebox(0,0)[cc]{$E_{1,2}^{k,-}$}}
\put(120.00,24.00){\makebox(0,0)[cc]{$E_{1,2}^{k,+}$}}
\put(97.00,15.00){\makebox(0,0)[cc]{$E_{2,2}^{k,-}$}}
\put(106.00,15.00){\makebox(0,0)[cc]{$E_{2,2}^{k,+}$}}
\put(37.00,15.00){\makebox(0,0)[cc]{$E_{2,1}^{k,-}$}}
\put(52.00,24.00){\makebox(0,0)[cc]{$E_{1,1}^{k,-}$}}
\put(67.00,24.00){\makebox(0,0)[cc]{$E_{1,1}^{k,+}$}}
\put(80.00,15.00){\makebox(0,0)[cc]{$E_{2,1}^{k,+}$}}
\put(125.00,15.00){\makebox(0,0)[cc]{$E_{2,3}^{k,-}$}}
\put(155.00,15.00){\makebox(0,0)[cc]{$E_{2,3}^{k,+}$}}
%\put(135.00,15.00){\makebox(0,0)[cc]{$E_{1,3}^{k,-}$}}
%\put(153.00,15.00){\makebox(0,0)[cc]{$E_{1,3}^{k,+}$}}
\put(134.00,24.00){\makebox(0,0)[cc]{$E_{1,3}^{k,-}$}}
\put(147.00,24.00){\makebox(0,0)[cc]{$E_{1,3}^{k,+}$}}
\end{picture}
%%%%%%%%%%%%%%%%%%%%%%%
\caption{Graph of the Lyapunov function $F_k(\l,q_\ve)$
for fixed $\ve\in(0,\ve_1)$.
The spectral bands of $H_k(q_\ve)$ are marked
by bold lines, counted with multiplicity. Here
$E_{\n,n}^{k,\pm}$ are endpoints of the bands:
$E_{1,3}^{k,\pm}=r_{k,2}^\pm$ and all other points
$E_{\n,n}^{k,\pm}$ satisfy
$F_{k,\n}(E_{\n,n}^{k,\pm})=(-1)^n$,
$E_{\n,2n}^{k,\pm}$ are periodic eigenvalues,
$E_{\n,2n+1}^{k,\pm}$ are antiperiodic eigenvalues.}
\lb{lfd}
\end{figure}
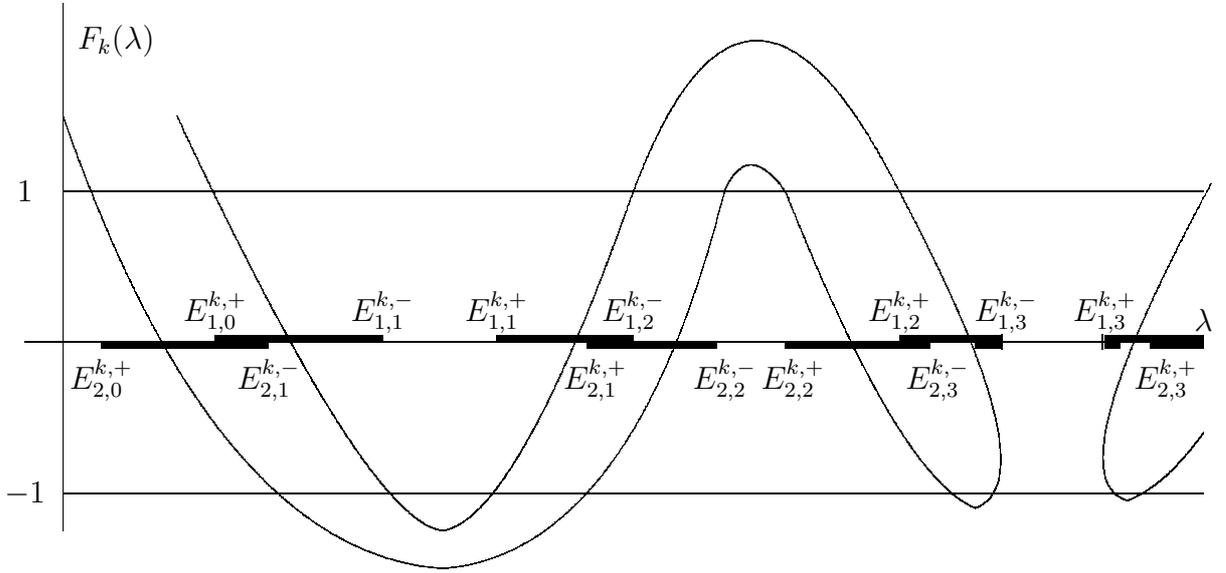

%%%%%%%%%%%%%

%%%%%%%%%%%

%%%%%%%%%%%%%

%%%%%%%%%%%%%%%

There are a lot of papers  about the spectral analysis of the
Schr\"odinger operator on periodic graph and periodic nanotubes.
Molchanov and Vainberg \cite{MV} consider
 Schr\"odinger operators with $q=0$ on so-called necklace  periodic graphs.
%  (which are similar to the simplest zigzag nanotobe, but the edges  of %the fundamental cell have the different lengths).
%In this case the Lyapunov function is meromorphic on the plane with %poles on the real line, which
% give the eigenvalues.
Korotyaev and Lobanov [KL], [KL1] consider the
Schr\"odinger operator on the zigzag nanotube.
The spectrum of this operator consists of an absolutely continuous part (intervals separated by gaps) plus an infinite number of eigenvalues  with infinite multiplicity. They describe all eigenfunctions with the same eigenvalue. They define a Lyapunov function,
which is analytic on some Riemann surface. On each
sheet, the Lyapunov function has the same properties
as in the scalar case, but it has
branch points (resonances). They prove that all  resonances are real and they determine the asymptotics of the periodic and anti-periodic spectrum and of the resonances at high energy. They show that there exist two types of gaps: i) stable gaps, where the endpoints are periodic and
anti-periodic eigenvalues, ii) unstable (resonance) gaps, where the
endpoints are resonances (i.e., real branch points of the Lyapunov
function).
They describe all finite gap potentials. They show that the mapping: potential $\to$ all eigenvalues is a real analytic isomorphism for some class of potentials.

Moreover, Korotyaev and Lobanov  [KL1] consider
magnetic Schr\"odinger operators on zigzag nanotubes.
They describe how the spectrum depends on the magnetic field.

 Korotyaev [K2] considers integrated density of states and effective masses  for zigzag nanotubes
in magnetic fields. He obtains a priori estimates of gap lengths in terms
of effective masses.

Kuchment and Post \cite{KuP} consider
the case of the zigzag, armchair  and achiral  nanotubes
with  $q\in L_{even}^2(0,1)$.
They show that the spectrum of the Schr\"odinger operator
(on these nanotubes), as a set, coincides with the spectrum  of the Hill
operator.
We would like to note that the case of even potential $q$
is simple and is closed to the case $q=0$.
Indeed, if $q$ is even, then  $F_-=0$ \cite{MW}
and by Theorem \ref{T3}, the corresponding Lyapunov functions
are expressed in terms of the Lyapunov function $F$
(for the Hill operator).
Moreover, using the identity $F(\l)=\cos \k(\l)$ ($\k(\l)$
is the quasimomentum for the Hill operator),
the Lyapunov function is expressed in terms of $\cos \k(\l)$,
which is similar to the case $q=0$, where $F(\l)=\cos\sqrt\l$.

%%%%%%%%%%%%
In contrast to \cite{KuP} we provide
spectral analysis for the Schr\"odinger operators with
arbitrary potentials on edges,
that can not be obtained using methods of \cite{KuP}.

%%%%%%%%%%%%%%%%%%%%%%%%%

%%%%%%%%%%%%%%

We present the plan of the paper.
In Sect. 2 we prove Theorem 1.2 about the eigenfunctions of $H_k$
and Theorem 1.3 about $\s_{ac}(H_k)$.
A technical proof of Theorem 1.1 is placed to Sect. 3.

\section{Spectrum of the operators $H_k$}
\setcounter{equation}{0}

\no {\bf Proof of Theorem \ref{T2}}. Proof of i) repeats the
arguments from \cite{KL}.

\no ii) Recall  $\vk_1=1-s^k\vp_1'^2$, $\vk_2=1-s^k\vp_1'^4$
and $\vp_1'\ne 0$.

If $k\not\in \{0,{N\/2}\}$, then $\Im s^k\ne 0$ which gives
$\Im\vk_1\ne 0,\Im\vk_2\ne 0$.

If $k={N\/2}$, then $s^k=-1$ and
$\vk_1=1+\vp_1'^2\ne 0,\vk_2=1+\vp_1'^4\ne 0$.

If $k=0$, then $s^k=1$ and we have  $\vk_1=1-\vp_1'^2,\vk_2=1-\vp_1'^4$.
Thus, we obtain

\no (a) If $\vp_1'^2\ne 1$ or $k\ne 0$, then $\vk_1\ne 0,\vk_2\ne 0$.

\no (b) If $\vp_1'^2=1$ and $k=0$, then $\vk_1=\vk_2=0$.

\no We will define the eigenfunctions of $H_k$: in
the case (a) $\p^{(0,1)}$ is given by
\[
\lb{T2-1}
\p^{(0,1)}_{n,j}=0, \ for\ all\ n\ne 0,1,\ j\in \N_6,\ \
and \
\p^{(0,1)}_{1,j}=0,\ j=5,6,\ \ \
\p^{(0,1)}_{0,5}=0,\ \ \p^{(0,1)}_{0,6}=\vk_2\vp,
$$
$$
s^k\vp_1'\p^{(0,1)}_{1,4}\!\!=\!\p^{(0,1)}_{1,1}\!\!=\!
{\p^{(0,1)}_{1,2}\/\vp_1'}\!=\!
{\p^{(0,1)}_{1,3}\/\vp_1'^2}\!=\!-s^k\vp_1'\p^{(0,1)}_{0,4}
\!\!=\!-\p^{(0,1)}_{0,1}\!\!=\!-{\p^{(0,1)}_{0,2}\/\vp_1'}
\!=\!-s^k\vp_1'^2\p^{(0,1)}_{0,3}
\!\!=\!s^k\vp_1'^2\vp,
\]
and the function $\p^{(0,2)}$ is given by
\[
\lb{T2-5}
\p^{(0,2)}_{n,j}=0, \ for\ all\ (n,j)\in (\Z\sm \{-1,0,1\})\ts\N_6,\ \
and \
\p^{(0,2)}_{-1,j}=0,\ j=1,3,4,
$$
$$
\p^{(0,2)}_{1,j}=0,\ j\ne 4,\ \ \
\p^{(0,2)}_{0,1}=0,\ \ \p^{(0,2)}_{0,3}=\vk_1\vp,
$$
$$
\p^{(0,2)}_{-1,2}=-\p^{(0,2)}_{-1,5}=s^k\p^{(0,2)}_{0,4}=
{\p^{(0,2)}_{-1,6}\/\vp_1'}=-\p^{(0,2)}_{0,2}
=\p^{(0,2)}_{0,5}=-s^k\p^{(0,2)}_{1,4}=
-\vp_1's^k\p^{(0,2)}_{0,6}=s^k\vp_1'\vp.
\]
In the case (b) the functions $\p^{(0,\n)},\n=1,2$ are given by
\[
\lb{T2-2}
\p^{(0,\n)}_{n,j}=0,\  (n,j)\in (\Z\sm\{ 0,1\})\ts\N_6,\qq
\p^{(0,\n)}_{1,j}=0, \ j\in \N_6\sm\{4\},
$$
$$
\p^{(0,\n)}_{1,4}=\vp, \qq \p^{(0,\n)}_{0,5}=-\vp,\qq
\p^{(0,\n)}_{0,6}=\vp_1'\vp,\qq
\p^{(0,1)}_{0,1}=\p^{(0,1)}_{0,3}=-\vp_1'\vp,
$$
$$
 \p^{(0,1)}_{0,2}=0,\qq
\p^{(0,1)}_{0,4}=-\vp,\qq
\p^{(0,2)}_{0,1}=\p^{(0,2)}_{0,3}=\p^{(0,2)}_{0,4}=0, \qq
\p^{(0,2)}_{0,2}=\vp.
\]

Using \er{T2-1}-\er{T2-2}, we deduce that  $\p^{(0,\n)}$ satisfy
the Kirchhoff conditions \er{1K0}, \er{1K1}. Thus
$\p^{(0,r)}$ are eigenfunctions  of $H_k$.
The operator $H_k$ is periodic, then each $\p^{(n,\n)}, (n,\n)\in \Z\ts
\N_2$ is an eigenfunction.
We will show that the sequence $\p^{(n,\n)}, (n,\n)\in \Z\ts
\N_2$ forms a basis for $\cH_k(\l)$.
The functions $\p^{(n,\n)}$ are linearly independent:

\no in the case (a) we have $\G_{n,6}\ss\supp\p^{(n,1)}\sm \supp\p^{(m,1)}$ and
$\G_{n,3}\ss\supp\p^{(n,2)}\sm \supp\p^{(m,2)}$ for all $n\ne m$
and $\G_{n,1}\ss\supp\p^{(n,1)}\sm \supp\p^{(n,2)}$ for all  $n\in\Z$;

\no in the case (b) we have $\cup_{j\in\N_6,j\ne 4}\G_{n,j}\ss\supp\p^{(n,\n)}\sm
\supp\p^{(m,\n)},\n=1,2,$
for  all  $n\ne m$, and $\G_{n,2}\ss\supp\p^{(n,2)}\sm \supp\p^{(n,1)}$ for  all  $n\in\Z$.

Consider the case (a). For any $f\in \cH_k(\l)$ we will show the identity
\er{T2-3}, i.e.,
\[
\lb{eit2-2}
f=\wh f, \ \  where \  \wh f=\!\!\!\!\sum_{(n,\n)\in \Z\ts
\N_2}\!\!\!\!\wh f_{n,\n}\p^{(n,\n)},\qq
\text{and}\qq \wh f_{n,1},\wh f_{n,2}\qq \text{are\ given\ by\ \er{T2-4}}.
\]
From $\l\in\s_D$, we deduce that $f|_{\G_{n,j}}=f_{n,j}'(0)\vp$.
The identity $f=\sum\wh f_{n,\n}\p^{(n,\n)}$
and \er{T2-1}, \er{T2-5} provide
$$
\wh f|_{\G_{n,6}}=\wh f_{n,1}\p^{(n,1)}|_{\G_{n,6}}+\wh
f_{n,2}\p^{(n,2)}|_{\G_{n,6}}
=(\wh f_{n,1}\vk_2-\wh f_{n,2})\vp,
$$
$$
\wh f|_{\G_{n,3}}=\wh f_{n,1}\p^{(n,1)}|_{\G_{n,3}}+\wh
f_{n,2}\p^{(n,2)}|_{\G_{n,3}}
=(-\wh f_{n,1}+\wh f_{n,2}\vk_1)\vp.
$$
Substituting \er{T2-4} into these identities we obtain
\[
\lb{idFf}
\wh f|_{\G_{n,6}}=f_{n,6}'(0)\vp=f|_{\G_{n,6}},\ \ \
\wh f|_{\G_{n,3}}=f_{n,3}'(0)\vp=f|_{\G_{n,3}} \ \ \ all \ \ n\in\Z.
\]
This yields
$\sum |\wh f_{n,\n}|^2<\iy $ and $\wh f\in L^2(\G^{(1)}).$

Note that $\wh f$ satisfies the Kirchhoff conditions \er{1K0}, \er{1K1}
and $-\wh f_\a''+q\wh f_\a=\l \wh f_\a, \a\in \cZ_1$.
Consider the function $u=f-\wh f$.
The function $u=0$ at all vertices of $\G^1$ and then
$u_{n,j}=C_{n,j}\vp, (n,j)\in\Z\ts\N_6$, $C_{n,3}=C_{n,6}=0$.
The Kirchhoff boundary conditions \er{1K0}-\er{1K1} yield $C_{n,2}=C_{n,4}=0,n\in\Z$. Assume that $C_{n,1}=C$. Then
$C_{n,5}={C_{n,1}\/\vp_1'}={C\/\vp_1'}$ and
$C_{n+1,1}={C_{n,5}\/\vp_1'}={C\/\vp_1'^n}$
for all $n\in\Z$. Since $u\in L^2(\G^1)$, we have $C=0$ and $u=0$,
which yields \er{eit2-2}.

Consider the case (b). For any $f\in \cH_k(\l)$ we will show the identity
\er{T2-3}, i.e.,
\[
\lb{T2c}
f=\wh f, \qq  where \ \ \wh f=\sum_{n\in \Z,r=1,2}\wh f_{n,r}\p^{(n,r)},
\qqq
\wh f_{n,2}=f_{n,2}'(0),\qqq
\wh f_{n,1}=-\vp_1' f_{n,1}'(0).
\]
From \er{T2c} and $\l\in \s_D$ we deduce that
\[
\lb{T2d}
\wh f|_{\G_{n,1}}=-\wh f_{n,1}\vp_1'\vp=f|_{\G_{n,1}},\qq
\wh f|_{\G_{n,2}}=\wh f_{n,2}\vp=f|_{\G_{n,2}}
 \ \ \ all \ \ n\in\Z.
\]
This yields
$\sum |\wh f_{n,\n}|^2<\iy$ and then $\wh f\in \cH(\l)$.

Consider the function $u=f-\wh f$.
The function $u=0$ at all vertices of $\G^1$,
which yields $u_{n,j}=C_{n,j}\vp,n\in\Z, j\in\Z_6,$ $C_{n,1}=C_{n,2}=0$.
The Kirchhoff boundary conditions
\er{1K0}-\er{1K1} yield $C_{n,5}=0$, and then $C_{n,4}=0,n\in\Z$.
Assume that $C_{n,3}=C$. Then
$C_{n,6}=-C_{0,3}=-C$ and $C_{n+1,3}=-C_{n,6}=C$ for all $n\in\Z$.
Due to $u\in L^2(\G^{(1)})$, we have $C=0$ and $u=0$, which yields
\er{T2c}.

The mapping $f\to (\wh f_{n,\n})_{(n,\n)\in \Z\ts\N_2}$ is a linear and one-to-one  mapping from $\cH_k$ onto $\ell^2\os \ell^2$.
Then it is a linear isomorphism. $\BBox$

\no {\bf Proof of Theorem \ref{T3}.} (i)
Using the standard arguments from [KL] for  $D_k(\cdot,\t)=\det(\cM_k-\t I_4)$ (see also \cite{BBK})
and \er{T1-2}, \er{T1-4}, \er{T1-3}
we obtain \er{S3b} and \er{DeLk}.

\no  The proof of (ii),(iii) repeats standard arguments from [KL].

\no (iv) If $q$ is even, then  $F_-=0$ (see p.8, \cite{MW}) and
identities \er{DeLk}  yield
\[
\lb{F0}
F_{0,1}={(3F+1)^2\/2}-1,\qq F_{0,2}={(3F-1)^2\/2}-1,
\]
\[
\lb{Fk0}
F_{k,1}={1\/2}\lt(\sqrt{9F^2-s_k^2}+|c_k|\rt)^2-1,\qq
F_{k,2}={1\/2}\lt(\sqrt{9F^2-s_k^2}-|c_k|\rt)^2-1.
\]
Identities \er{F0} show that
$\{\l\in\R: F_{0,\nu}(\l)\in [-1,1]\ \text{for some}\
\nu\in\N_2\}=\{\l\in\R:F(\l)\in [-1,1]\}$,
which together with \er{T3-1} yields $\s_{ac}(H_0)=\s(\wt H)$.
Identities \er{T3-1}, \er{Fk0} provide
$$
\s_{ac}(H_k)=\{\l\in\R: F_{k,\nu}(\l)\in [-1,1]\ \text{for some}\
\nu\in\N_2\}\ss
$$
$$
\ss\{\l\in\R: F_{0,\nu}(\l)\in [-1,1]\ \text{for some}\
\nu\in\N_2\}=\s_{ac}(H_0)
$$
for all $k\in\Z_N$, which yields
$\s_{ac}(H_0)=\s_{ac}(H)$.$\BBox$

\no {\bf Proof of Proposition \ref{P7}.}
If $q_\ve={1\/\ve}\d(t-a),\ve\ne 0,
a\in (0,1)$, then we have
$$
\vt(t,\l,q_\ve)=\cos z t+\cos z a{\sin z (t-a)\/\ve z},\qq
\vp(t,\l,q_\ve)={\sin z t\/z}+{\sin z a\/\ve z}{\sin z (t-a)\/z},
$$
$$
F(\l,q_\ve)=\cos z +{\sin z\/2z\ve},\qq
F_-(\l,q_\ve)={\sin z(2a-1)\/2z\ve},\qq z=\sqrt\l.
$$
Let $a={1\/2}+{c_k}\ve+{\ve^2},k\not\in\{0,{N\/2}\}$. Then we obtain
\[
\lb{P4b}
F_-(\l,q_\ve)={\sin 2z({c_k \ve}+{\ve^2})\/2z\ve}={c_k}+{\ve}+O(\ve^{2}),\qq
F_-^2(\l,q_\ve)-c_k^2={2c_k\ve}+O(\ve^{2})
\]
as $|\ve|\to 0$, uniformly on $|z|\le \pi n_0$.
Using the identities \er{DeLk}, we rewrite the equation $\r_k(\l,q_\ve)=0,
\l\in \C$
in the form
$
F_-^2(\l,q_\ve)-c_k^2=-{9F^2(\l,q_\ve)c_k^2\/s_k^2},
$
and finally
%We rewrite this equation in the form
\[
\lb{Pre}
\F_+(\l,\ve)\F_-(\l,\ve)=0,\qq \F_\pm(\l,\ve)=\ve\lt({3F(\l,q_\ve)c_k\/s_k}\pm
i\sqrt{F_-^2(\l,q_\ve)-c_k^2}\rt), \qqq \l\in \C.
\]
Using $\F_{\pm}(\l,0)={c_k\sin z\/2s_k z}$ and
${\pa\F_\pm\/\pa\l}((\pi n)^2,0)\ne 0,n\ge 1$,
and applying the Implicit Function Theorem to the functions $\F_\pm$
we conclude that
for each $n\le n_0$ there exists $\ve_n>0$ such that equation \er{Pre}
has exactly
two roots $r_n^\pm(\ve),r_n^\pm(0)=(\pi n)^2$, in the disc $|\ve|<\ve_n$.

Thus we have $\sqrt{r_n^\pm(\ve)}=\pi n+\t_n^\pm,\t=\t_n^\pm\to 0$
as $|\ve|\to 0$
and then
\[
\lb{P4c}
F(r_n^\pm(\ve),q_\ve)=(-1)^n\lt(\cos \t +{\sin \t\/2(\pi
n+\t)\ve}\rt)={(-1)^n\/2\pi n\ve}(\ve+\t+O(\t^2)).
\]
Substituting \er{P4b}, \er{P4c} into $
F_-^2(\l,q_\ve)-c_k^2=-{9F^2(\l,q_\ve)c_k^2\/s_k^2},
$ we obtain
\[
\lb{P4e}
{2c_k\ve}+O(\ve^{2})=-{9c_k^2\/s_k^2}\lt({\ve+\t+O(\t^2)\/2\pi n\ve}\rt)^2,
\qq \text{and\ then}\qq \t=O(\ve).
\]
Using \er{P4e} again, we have
$
\t=-\ve\pm{2\pi n\ve^{3/2}s_k\/3c_k}\sqrt{-{2c_k}}+O(\ve^{2}),
$
which yields \er{P4-1}. $\BBox$

%%%%%%%%%%%%%%%%%%%%%%%%%%%%%%%%%%%%%%%%%%%%

\section{ Fundamental solutions}
\setcounter{equation}{0}

\no {\bf Proof of Theorem \ref{T1}.}
i) Define the operator $\cS$ in $\C^N$ by $\cS u=(u_N,u_1,\dots,u_{N-1})^\top$, $u=(u_n)_1^N\in \C^N$.
The unitary operator $\cS$ has the form $\cS =\sum_1^Ns^k\cP_k$, where $\cS e_k=s^ke_k$ and
$e_k={1\/N^{1\/2}}(1,s^{-k},s^{-2k},...,s^{-kN+k})$
 is an eigenvector (recall $s=e^{i{2\pi \/N}}$);
 $\cP_ku=e_k(u,e_k)$ is a projector. The function $f$ in the Kirchhoff boundary conditions \er{KirC}
is a vector function  $f=(f_{\o}), \o=(n,j,k)\in\cZ$.  We define a new
vector-valued function $f_{n,j}=(f_{n,j,k})_{k=1}^{N}\in \C^N, (n,j)\in
\cZ_1=\Z\ts \N_6$, which  satisfies the equation
$-f_{n,j}''+qf_{n,j}=\l f_{n,j}$,
and the conditions
for all $n\in Z$, which follow from the Kirchhoff conditions \er{KirC}:
\begin{multline}
\label{C0}
f_{n,1}(1)=f_{n,2}(0)=f_{n,5}(0),\quad
f_{n,2}(1)=f_{n,3}(0)=f_{n,6}(0),\\
f_{n,3}(1)=f_{n,4}(0)=f_{n-1,6}(1),\quad
\cS f_{n,4}(1)=f_{n,1}(0)=f_{n-1,5}(1),
\end{multline}
\begin{multline}
\label{C1}
f_{n,1}'(1)-f_{n,2}'(0)-f_{n,5}'(0)=0,\qq
f_{n,2}'(1)-f_{n,3}'(0)-f_{n,6}'(0)=0,\\
f_{n,3}'(1)-f_{n,4}'(0)+f_{n-1,6}'(1)=0,\qq
\cS f_{n,4}'(1)-f_{n,1}'(0)+f_{n-1,5}'(1)=0.
\end{multline}
The operators $\cS$ and $\mH$ commute, then  we deduce that $\mH\cP_k$ is unitarily equivalent to the operator $H_k$ acting
 in $L^2(\G^1)$ and $H_k$ is given by
$ (H_kf)_\a=-f_\a''+q(t)f_\a$, where $(f_\a)_{\a\in \cZ_1} ,(f_\a'')_{\a\in \cZ_1}\in L^2(\G^1)$ and components $f_\a$
satisfy the boundary conditions \er{1K0}, \er{1K1}.
Thus $\mH$ is unitarily equivalent to the operator $H=\os_1^N H_k$

\no ii) We prove the following idenities:
\[
\lb{T11}
\cM_k=\cR^{-1}Y\cT_k \cR,\qq
\cT_k=\ma V_k & I_2 \\ V_0V_k-I_2 & V_0 \am,\ \
\ \
V_k=\ma 2F & -s^k\\ -s^{-k} & 2F\am,\ \ Y=\ma\vt_1 I_2& I_2\\
\vp_1\vt_1' I_2&\vp_1' I_2\am.
\]
Let $\vp_t=\vp(t,\l),\vp_t'=\vp'(t,\l),...$
Recall that any solution $y$  of the equation
$-y''+qy=\l y$ on $[0,1]$ satisfies
\[
\lb{T1a}
y(t)=y(0)\vt_t+{\vp_t\/\vp_1}(y(1)-\vt_1y(0)),\qq
\ma y(1)\\y'(1)\am=\cM \ma y(0)\\y'(0)\am,\qq
\cM^{-1}=\ma\vp_1'&-\vp_1\\-\vt_1'&\vt_1\am.
\]

Let $\cF_{n}(t)=(f_{n,5}(t),f_{n,6}(t),f_{n,5}'(t),f_{n,6}'(t))^\top $.
We will prove that for each vector
$\cF_{-1}(1)=h=(h_5,h_6,h_5',h_6')^\top\in\C^4$ there
exists a unique vector function $(f_\a)_{\a\in\cZ_1}$,
satisfying the equations $-f_\a''+q(x)f_\a=\l f_\a,x\in[0,1],\a\in\cZ_1$,
and the Kirchhoff boundary conditions \er{1K0}, \er{1K1}.

{\bf Firstly}, we will determine $f_{0,1},f_{0,3}$ in terms of $h$.
The Kirchhof conditions \er{1K0}, \er{1K1} at $n=0$ yield
\begin{multline}
\lb{T1v}
f_{0,3}(1)=f_{0,4}(0)=h_6,\qq
s^k f_{0,4}(1)=f_{0,1}(0)=h_5,\\
f_{0,3}'(1)-f_{0,4}'(0)+h_{6}'=0,\qq
s^kf_{0,4}'(1)-f_{0,1}'(0)+h_{5}'=0.
\end{multline}
Let $w_{13}(t)=(f_{0,1}(t),f_{0,3}(1-t), f_{0,1}'(t), f_{0,3}'(1-t))^\top $.
Then conditions \er{T1v} imply
%$\cF_{-1}(1)=(f_{-1,5}(1),f_{-1,6}(1),f_{-1,5}'(1),f_{-1,6}'(1))^\top $,
\[
\lb{T1z1}
w_{13}(0)=\ma I_2& 0\\ 0& {\bf j}_1\am h+w_4,\qq
w_4=(0,0,s^kf_{0,4}'(1),f_{0,4}'(0))^\top,\qq
{\bf j}_1=\ma 1& 0\\ 0 & -1\am.
\]
Using \er{T1a}, \er{T1v} we get
$
f_{0,4}(t)=h_6\vt_t+{\vp_t\/\vp_1}(s^{-k}h_5-\vt_1h_6), \qq t\in[0,1],
$ which yields
$$
f_{0,4}'(0)={1\/\vp_1}(s^{-k}h_5-\vt_1h_6),
\qq
f_{0,4}'(1)=h_6\vt_1'+{\vp_1'\/\vp_1}(s^{-k}h_5-\vt_1h_6)
={1\/\vp_1}(s^{-k}\vp_1'h_5-h_6),
$$
where we have used the identity $\vt_1\vp_1'-\vt_1'\vp_1=1$.
The last identities and \er{T1z1} give
\[
\lb{T1u}
w_4=\ma 0 & 0\\ {A_0\/\vp_1} & 0 \am h,\qq
A_0=\ma\vp_1' & -s^k \\ s^{-k}& -\vt_1\am,\qqq
w_{13}(0)=Y_0 h,\qq Y_0=\ma I_2& 0\\ {A_0\/\vp_1}&{\bf j}_1\am.
\]
Thus for each $h=\cF_{-1}(1)\in\C^4$ there exists unique
$w_{13}(0)=(f_{0,1}(0),f_{0,3}(1), f_{0,1}'(0), f_{0,3}'(1))^\top $ and then the unique functions
$f_{0,1}(t),f_{0,3}(t), t\in [0,1]$ satisfying the equation $-f''+qf=\l f$
and the Kirchhoff boundary conditions \er{T1v}.

Identities \er{T1a} imply
%again we see that these functions satisfy
%\[\lb{T1b}
$\ma f_{0,1}(1)\\ f_{0,1}'(1)\am
=\cM \ma f_{0,1}(0)\\ f_{0,1}'(0)\am,\ \
\ma f_{0,3}(0)\\ f_{0,3}'(0)\am
=\cM^{-1} \ma f_{0,3}(1)\\ f_{0,3}'(1)\am,
$
%\]
which yields
\[
\lb{T1t}
w_{13}(1)= Y_1w_{13}(0),\qqq Y_1=\ma A_1{\bf j}_2 & \vp_1{\bf j}_1\\
\vt_1'{\bf j}_1 & {\bf j}_2 A_1\am,\qq
A_1=\ma 0& \vt_1\\ \vp_1' & 0\am,\qq {\bf j}_2 =\ma 0& 1\\ 1& 0\am.
\]

{\bf Secondly,} we will determine $f_{0,5},f_{0,6}$ in terms of $f_{0,1},f_{0,3}$. The Kirchhof conditions \er{1K0}, \er{1K1} at $n=0$ yield
\[
\lb{T1z}
f_{0,1}(1)=f_{0,2}(0)=f_{0,5}(0),\quad
f_{0,2}(1)=f_{0,3}(0)=f_{0,6}(0),
$$
$$
f_{0,1}'(1)-f_{0,2}'(0)-f_{0,5}'(0)=0,\qq
f_{0,2}'(1)-f_{0,3}'(0)-f_{0,6}'(0)=0,
\]
which gives
\[
\lb{T1y}
\cF_0(0)=\ma I_2& 0\\ 0& {\bf j}_1\am w_{13}(1)+w_2,\qq
w_2=(0,0,-f_{0,2}'(0),f_{0,2}'(1))^\top ,
\]
where $w_{13}(1)=(f_{0,1}(1),f_{0,3}(0), f_{0,1}'(1), f_{0,3}'(0))^\top $,
$\cF_{0}(0)=(f_{0,5}(0),f_{0,6}(0),f_{0,5}'(0),f_{0,6}'(0))^\top $.
Using the Kirchhof conditions \er{T1z}
and \er{T1a}, we obtain
$$
f_{0,2}(t)=f_{0,1}(1)\vt_t+{\vp_t\/\vp_1}(f_{0,3}(0)-\vt_1f_{0,1}(1)),\qq
%$$$$
f_{0,2}'(0)={1\/\vp_1}(f_{0,3}(0)-\vt_1f_{0,1}(1)),
$$
$$
f_{0,2}'(1)=f_{0,1}(1)\vt_1'+{\vp_1'\/\vp_1}(f_{0,3}(0)-\vt_1f_{0,1}(1))
={1\/\vp_1}(\vp_1'f_{0,3}(0)-f_{0,1}(1)),
$$
where we have used the Wronskian $\vt_1\vp_1'-\vt_1'\vp_1=1$.
Then
\[
\lb{a3}
w_2=\ma 0 & 0\\ {1\/\vp_1}A_2 &0 \am w_{13}(1),\qqq
A_2=\ma \vt_1 & -1 \\ -1 & \vp_1'\am.
\]
Substituting  this identity into \er{T1y} we obtain
\[
\lb{T1s}
\cF_{0}(0)=Y_2 u(1),\qqq Y_2=\ma I_2& 0\\{1\/\vp_1}A_2& {\bf j}_1\am.
\]
Thus for each $h\in\C^4$ there exists a unique vector
$\cF_0(0)=(f_{0,5}(0),f_{0,6}(0), f_{0,5}'(0), f_{0,6}'(0))^\top $
and the unique functions $f_{0,5}(t),f_{0,6}(t), t\in[0,1]$ satisfying the equation $-f''+qf=\l f$
and the Kirchhoff boundary conditions \er{T1z}.
Identities \er{T1a} give
$\ma f_{0,5}(1)\\ f_{0,5}'(1)\am
=\cM \ma f_{0,5}(0)\\ f_{0,5}'(0)\am $ and $\ma f_{0,6}(1)\\ f_{0,6}'(1)\am
=\cM \ma f_{0,6}(0)\\ f_{0,6}'(0)\am
$, then
\[
\lb{T1w}
\cF_0(1)=Y_3 \cF_0(0),\qqq Y_3=\ma\vt_1 I_2&\vp_1 I_2\\ \vt_1' I_2&\vp_1'
I_2\am.
\]

Thus we have proved that for each $h\in\C^4$ there exist the unique functions
$f_{0,j}(t),j\in\N_6$, satisfying the equation $-f''+qf=\l f$
and the Kirchhoff boundary conditions \er{1K0}, \er{1K1}.
By the periodicity,
we obtain similar results for all functions $f_{\a},\a\in \cZ_1$.

Recall that the fundamental solutions
$\vt_k^{(\n)}=(\vt_{k,\a}^{(\n)})_{\a\in \cZ_1},
\vp_k^{(\n)}=(\vp_{k,\a}^{(\n)})_{\a\in \cZ_1},\n=1,2$
 satisfy \er{eqf0} and the monodromy matrix $\cM_k$ is given by \er{dmm}.
%\[
%\cM_k=\ma \vT_{k,0}(1,\l) & \F_{k,0}(1,\l)\\
%\vT_{k,0}'(1,\l) & \F_{k,0}'(1,\l)\am\!\!,\ \
%\vT_{k,n}=\ma\vt_{k,n,5}^{(1)} & \vt_{k,n,5}^{(2)} \\
%\vt_{k,n,6}^{(1)} & \vt_{k,n,6}^{(2)}\am,\ \
%\F_{k,n}=\ma\vp_{k,n,5}^{(1)} & \vp_{k,n,5}^{(2)} \\
%\vp_{k,n,6}^{(1)} & \vp_{k,n,6}^{(2)}\am.
%\]
Then $\cF_{0}(1)=\cM_k h,\ h=\cF_{-1}(1),$
and identities \er{T1u}, \er{T1t}, \er{T1s}, \er{T1w} yield
\[
\lb{T1r}
\cM_k=Y_3Y_2Y_1 Y_0.
\]

Now we calculate $Y_2Y_1 Y_0$.
We need the identities
\[
\lb{a1}
\vp_1'+\vt_1=2F,\qqq \vt_1'\vp_1+\vp_1'^2=2\vp_1'F-1,\qqq
\vt_1'\vp_1+\vt_1^2=2\vt_1F-1.
\]
Identities \er{T1u}, \er{T1t} give
\[
\lb{T1q}
Y_1Y_0=\ma A_1{\bf j}_2 +{\bf j}_1A_0&  \vp_1 I_2\\
\vt_1'{\bf j}_1+{1\/\vp_1}{\bf j}_2 A_1A_0&
{\bf j}_2 A_1{\bf j}_1\am
=\ma V_k&  \vp_1 I_2 \\ {1\/\vp_1}B_1 & B_2\am,\qqq
B_2=\ma\vp_1'&0\\0&-\vt_1\am,
\]
\[
\lb{a2}
B_1=\ma \vp_1\vt_1'& 0\\ 0 & -\vp_1\vt_1'\am
+\ma \vp_1'& 0\\ 0& \vt_1\am\ma\vp_1' & -s^k \\ s^{-k}&
-\vt_1\am
=\ma {2\vp_1'F-1}&-s^k{\vp_1'}\\s^{-k}{\vt_1}&1-{2\vt_1F}\am,
\]
where we have used \er{a1} and $V_k$ are given by \er{T11}.
Identities \er{T1s}, \er{T1q} imply
\[
\lb{T1x}
Y_2Y_1 Y_0
=\ma I_2& 0\\{1\/\vp_1}A_2&{\bf j}_1\am
\ma V_k& \vp_1 I_2\\ {1\/\vp_1}B_1 & B_2\am
=\ma V_k& \vp_1
I_2\\{1\/\vp_1}(A_2V_k+{\bf j}_1B_1)&A_2+{\bf j}_1B_2\am=\cR^{-1}\cT_k \cR,
\]
where
\[
\lb{a4}
\cT_k=\ma V_k&  I_2\\A_2V_k+{\bf j}_1B_1&A_2+{\bf j}_1B_2\am.
\]
Using \er{a3}, \er{T1q}, \er{a2} we have
$$
A_2V_k+{\bf j}_1B_1=\ma \vt_1 & -1 \\ -1 & \vp_1'\am\ma 2F & -s^k\\
-s^{-k} & 2F\am
+\ma 1& 0\\ 0 & -1\am\ma
{2\vp_1'F-1}&-s^k{\vp_1'}\\s^{-k}{\vt_1}&-{2\vt_1F+1}\am
=V_0V_k-I_2,
$$
$$
A_2+{\bf j}_1B_2=\ma \vt_1 & -1 \\ -1 & \vp_1'\am+\ma 1& 0\\ 0 &
-1\am\ma\vp_1'&0\\0&-\vt_1\am
=V_0.
$$
Substituting the last two identities into \er{a4}, we obtain the identity for
$\cT_k$ in \er{T11}.
Substituting  identity \er{T1x} into \er{T1r},
we obtain $\cM_k=Y_3\cR^{-1}\cT_k\cR$
which yields \er{T11}.
Identities \er{T11} show that the matrix-valued function
$\cR\cM_k\cR^{-1}$ is entire.

Identities \er{T1u}, \er{T1s} give $\det Y_0=\det Y_2=-1$. Identity \er{T1t}
shows that $Y_1$ is obtained
from the matrix $\ma\cM&0\\ 0&\cM\am$ by the transposition of rows and
columns, hence $\det Y_1=1$.
\er{T1w} implies that $Y_1$ is obtained
from the matrix $\ma\cM&0\\ 0&\cM^{-1}\am$ by the transposition of rows
and columns,
hence $\det Y_3=\det \cM\det \cM^{-1}=1$. Then \er{T1r} gives $\det\cM_k=1.$

Furthermore, \er{T11} gives
$$
\Tr\cM_k=\Tr\ma\vt_1I_2 & \vp_1I_2\\ \vt_1'I_2 & \vp_1'I_2\am
\ma V_k& \vp_1 I_2\\ {1\/\vp_1}(V_0V_k-I_2) & V_0\am
=\Tr(\vt_1V_k+V_0V_k-I_2+\vp_1\vt_1'I_2+\vp_1'V_0),
$$
\[
\lb{T1m}
\Tr V_k=\Tr V_0=4F,\qq \Tr (V_0V_k)=8F^2+s^{-k}+s^k.
\]
Hence
$$
\Tr
\cM_k=4F(\vt_1+\vp_1')+8F^2+s^{-k}+s^k-2+2\vp_1\vt_1'
=16F^2+s^{-k}+s^k-2+2\vp_1\vt_1'
=\Tr \cM_0-4s_k^2,
$$
where $\Tr \cM_0=16F^2+2\vp_1\vt_1'$, which yields \er{T1-2}.

We prove \er{T1-4}. We have
\[
\lb{a5}
\Tr\cM_k^2=\Tr(\vt_1V_k+V_0V_k-I_2)^2
+2\Tr
(\vt_1I_2+V_0)(\vp_1\vt_1'V_k+\vp_1'(V_0V_k-I_2))+\Tr(\vp_1\vt_1'I_2+\vp_1'V_0)^2.
\]
Direct calculations give
\[
\lb{a6}
\Tr V_k^2=\Tr V_0^2=8F^2+2,\qq
\Tr (V_0V_k)^2
=32F^4+16F^2(s^k+s^{-k}+1)+s^{2k}+s^{-2k},
\]
\[
\lb{a7}
\Tr(\vt_1V_k^2V_0+\vp_1'V_0^2V_k)=32F^4+8F^2(s^k+s^{-k}+1).
\]
Substituting identities \er{T1m}, \er{a6} and \er{a7} into \er{a5}
we obtain
$$
\Tr\cM_k^2\!\!=\!\!128F^4+8F^2+32F^2\vp_1\vt_1'+(32F^2+4\vp_1\vt_1')(s^k+s^{-k})
+s^{2k}+s^{-2k}-8\vt_1\vp_1'+2\vp_1^2\vt_1'^2+2
$$
$$
={1\/2}(\Tr\cM_0)^2+72F^2-8s_k^2\Tr \cM_0
-4s_{2k}^2-4=\Tr\cM_0^2-8s_k^2\Tr\cM_0-4s_{2k}^2,
$$
where we have used the identities
$s^k+s^{-k}=2-4s_k^2,\vt_1\vp_1'=\vp_1\vt_1'+1$.
This yields \er{T1-4}.

Now we will prove \er{T1-3}. Direct calculation show that $Y^\top JY=J$
and
$$
\cT_k^\top J\cT_k=\ma V_k^\top {\bf j}_2 (V_0V_k-I_2)-(V_0V_k-I_2)^\top {\bf j}_2 V_k
& V_k^\top {\bf j}_2 V_0-(V_0V_k-I_2)^\top  {\bf j}_2 \\
{\bf j}_2 (V_0V_k-I_2)-V_0{\bf j}_2 V_k & {\bf j}_2 V_0-V_0{\bf j}_2 \am=J,
$$
where the identities ${\bf j}_2 V_k=V_k^\top {\bf j}_2 ,{\bf j}_2 V_0=V_0{\bf j}_2 $
was used. Then $(Y\cT_k)^\top  JY\cT_k=J$.
Using the identities
$\cR J\cR=\vp_1 J,\ \cR^{-1}J\cR^{-1}={1\/\vp_1}J,$
we have  \er{T1-3}.
$\BBox$

 \no {\bf Acknowledgments.} 
Evgeny Korotyaev was partly supported by DFG project BR691/23-1.
The various parts of this paper were written at  ESI, Vienna , E. Korotyaev is grateful to the Institute for the hospitality.
A. Badanin is grateful to the Mathematical Institute of Humboldt Univ. for the hospitality.

\end{document}